\documentclass[12pt] {article}
\usepackage{jheppub}
\usepackage{bm}
\title{Conservative entropic forces}
\author{Matt Visser}
\affiliation{School of Mathematics, Statistics, and Operations Research, \\
Victoria University of Wellington, PO Box 600, Wellington 6140, New Zealand}
\emailAdd{matt.visser@msor.vuw.ac.nz}
\abstract{
Entropic forces have recently attracted considerable attention as ways to reformulate, retrodict, and perhaps even ``explain'' classical Newtonian gravity from a rather specific thermodynamic perspective. In this article I point out that if one wishes to reformulate classical Newtonian gravity in terms of an entropic force, then the fact that Newtonian gravity is described by a conservative force places significant constraints on the form of the entropy and temperature functions. (These constraints also apply to entropic reinterpretations of electromagnetism, and indeed to any conservative force derivable from a potential.) 

The constraints I will establish are sufficient to present real and significant problems for any reasonable variant of Verlinde's entropic gravity proposal, though for technical reasons the constraints established herein do not directly impact on either Jacobson's or Padmanabhan's versions of entropic gravity. In an attempt to resolve these issues, I will extend the usual notion of entropic force to multiple heat baths with multiple ``temperatures'' and multiple ``entropies''. 

\bigskip
\noindent
14 September 2011; 20 October 2011; \LaTeX-ed \today.

}
\begin{document}
\maketitle
\newpage
\section{Introduction} \label{sec: intro}
\def\d{{\mathrm{d}}}
\def\r{{\bm{r}}}
\def\a{{\bm{a}}}
\def\n{{\bm{\hat n}}}
\def\F{{\bm{F}}}
\def\bnabla{{\bm{\nabla}}}

Thermodynamic interpretations (and reinterpretations) of the Einstein equations of general relativity have been investigated for quite some time. 
Fundamental papers date back to Jacobson's work in 1995~\cite{Jacobson}, with closely related follow up articles appearing in references~\cite{Jacobson:1999, Jacobson-Parentani, Eling, Eling:2008, Chirco}.
Another significant thread to the relativistic discussion of thermodynamic interpretations (and reinterpretations) of the Einstein equations has been that due to Padmanabhan~\cite{Padmanabhan}, and his collaborators~\cite{Paranjape,  Mukhopadhyay, Kothawala, Kothawala:2009, Kolekar}, in the first decade of the 2000's.  Further afield, Hu and his collaborators have sought to interpret general relativity in a hydrodynamic~\cite{Hu:2002} and stochastic~\cite{Hu:1999, Hu-Verdaguer} manner. Entropic and thermodynamic issues also implicitly underly much of Sakharov's ``induced gravity'' approach~\cite{Sakharov, Visser:Sakharov}, and much of the ``analogue spacetime'' programme~\cite{Barcelo, Visser:ergo}.  For selected further developments see references~\cite{Sotiriou:2006, Sotiriou:2007, Liberati:2009, Sindoni:2009, Sindoni:2010, Chirco:2010}.

Set against this backdrop, in early 2010 Erik~Verlinde introduced a specific model for the interpretation of Newtonian gravity as an entropic force~\cite{Verlinde}; a model that has attracted considerable attention. Attempts have been made to extend Verlinde's model to loop quantum gravity~\cite{Smolin}, to the Coulomb force~\cite{Wang}, to Yang--Mills gauge fields~\cite{Freund}, to non-commutative geometries~\cite{Nicolini}, and perhaps more controversially to cosmology~\cite{Cai, Li, Easson}.  (For  earlier related work see~\cite{Cai:2005, Cai:2007}.) There have also been a number significant technical criticisms of Verlinde's ideas~\cite{Hossenfelder, Kobakhidze, Gao, Hu:2010, Kobakhidze:2011}. 

In the current article I shall not attempt to ``derive'' or ``justify'' an entropic interpretation for Newtonian gravity, rather I shall ask the converse question: ``\emph{Assuming that Newtonian gravity can be described by an entropic force, what does this tell us about the relevant temperature and entropy functions of the assumed thermodynamic system?}'' Starting from the definition of an entropic force
\begin{equation}
\F =  T \; \bnabla S,
\end{equation}
and asking that this entropic force reproduces the conservative force law of Newtonian gravity
\begin{equation}
\F = -\bnabla\Phi,
\end{equation}
one can deduce some rather strong constraints on the functional form of the temperature and entropy functions. (I specifically use $\Phi$ rather than $V$ for the potential to avoid any possibility of confusion with the notion of ``thermodynamic volume''.) 
\emph{The constraints we shall derive are strong enough to cause considerable unease and discomfort for the most direct implementation of Verlinde's ideas}, though I shall demonstrate that there are (somewhat strained) modifications of Verlinde's specific entropic force scenario that can be made to accurately reproduce Newtonian gravity. 

The issues I raise in this article do not directly affect the entropic/thermodynamic considerations of the fully relativistic models introduced by either Jacobson~\cite{Jacobson, Jacobson:1999, Jacobson-Parentani, Eling, Eling:2008, Chirco} or Padmanabhan~\cite{Padmanabhan, Paranjape,  Mukhopadhyay, Kothawala, Kothawala:2009, Kolekar};  those models should be assessed on their own merits.  It should be emphasised that Verlinde's scenario~\cite{Verlinde} makes considerably stronger assertions than either the Jacobson or Padmanabhan scenarios. 
The differences between the Jacobson and Padmanabhan scenarios and Verlinde's proposal are both technical and conceptual:
\begin{itemize}
\item 
Because the Jacobson and Padmanabhan scenarios are fully relativistic and fully field theoretic, the ultimate goal of those proposals is to obtain the Einstein equations, essentially as consistency conditions on the thermodynamics of an infinite collection of arbitrary observer-dependent virtual causal horizons. The focus there is on the field equations in some fiducial region of spacetime, not on any individual particle subject to buffeting by real physical thermal fluctuations. In those scenarios there is no direct statement or claim regarding the existence of  any $\F =  T \; \bnabla S$ physical entropic force, nor indeed any direct way of extracting any such force from their formalism.

\item
Because the Jacobson and Padmanabhan scenarios seek to derive the Einstein field equations, (or sometimes modified Einstein equations if one adopts non-standard thermodynamics for the virtual causal horizons), not individual forces on individual particles, there is no direct connection to the Verlinde scenario. Rather indirectly, one could appeal to the  Einstein--Infeld argument whereby the Einstein field equations applied to an isolated lump of stress-energy lead to the geodesic equations, and then further make weak-field assumptions to obtain Newtonian gravity via the geodesic equations --- but the logical distance between the thermodynamic aspects of the  Jacobson and Padmanabhan scenarios and Newton's inverse square law is rather significant. 

\item 
In direct contrast, in Verlinde's scenario (Newtonian) gravity is very literally to be interpreted as a $\F =  T \; \bnabla S$ physical entropic force, with the implication of a strong thermal coupling to some sort of physical heat bath. No such assumptions are made in either the Jacobson or Padmanabhan scenarios --- insofar as their proposals contain heat baths, they are virtual heat baths, each being attached to an observer-dependent virtual causal horizon.

\item
Furthermore Verlinde also makes a number of rather strong assertions regarding ``holographic screens'' that have no clear counterpart in either the Jacobson or Padmanabhan scenarios. In particular the ``holographic screens'' are typically taken to lie on equipotential surfaces, while no such limitation is imposed on the virtual causal horizons in either the Jacobson or Padmanabhan scenarios --- and the \emph{absence} of any such limitation is essential in turning integral constraints  evaluated on causal horizons into local field theoretic constraints (the Einstein equations) on the spacetime geometry. 

\item
In addition, a purely technical difference between the scenarios is that the holographic screens are timelike while the observer-dependent virtual causal horizons are null (lightlike). 
\end{itemize}
In summary, the  Jacobson or Padmanabhan scenarios are sufficiently different in their specific details from Verlinde's proposal that 
no direct conclusions regarding the former should necessarily be drawn from any problems encountered by the latter (or vice versa).

\section{Conservative entropic forces} \label{S: entropic}

There is no doubt that entropic forces exist, there are numerous physical examples, the most well-known of which are:
\begin{itemize}
\item elasticity of a freely jointed polymer;
\item hydrophobic forces;
\item osmotic forces;
\item colloidal suspensions;
\item binary hard sphere mixtures;
\item molecular crowding/depletion forces.
\end{itemize}
For instance, the configurational entropy of a freely jointed polymer immersed in a heat bath leads to an approximate Hooke's law relationship for the force required to hold the endpoints some fixed distance apart --- and I emphasise that this is a completely reversible force~\cite{Muller, Smith}. 
The question at hand, however,  is whether entropic forces can be used to mimic Newtonian gravity, or more generally any conservative force derivable from a potential, \emph{and whether this can be done in a manner consistent with Verlinde's specific proposal}. For definiteness we shall focus on two specific cases:
\begin{itemize}
\item A single particle interacting with an externally specified potential. Physical quantities are then dependent on a \emph{single} position variable $\r$.
\item A many-body system of $n$ mutually interacting particles. Physical quantities are then dependent on $n$ position variables $\r_i$, for $i\in\{1\dots n\}$.
\end{itemize}

\subsection{$1$-body external potential scenario} \label{SS: 1-body}

When a single body is interacting with an externally specified potential the force on the particle is
\begin{equation}
\F(\r) = -\bnabla\Phi(\r).
\end{equation}
If we \emph{assume} this can be mimicked by an entropic force then we must have
\begin{equation}
\F(\r) = T(\r) \; \bnabla S(\r),
\end{equation}
implying
\begin{equation}
\bnabla\Phi(\r) = - T(\r) \; \bnabla S(\r).
\label{E: key-1}
\end{equation}
Without any calculation, since $\bnabla \Phi || \bnabla S$, this immediately implies that the level sets of the potential are also level sets of the entropy. (Thus implying that the entropy is some function of the potential.) But by taking the curl of equation (\ref{E: key-1}) we also see $\bnabla T || \bnabla S$, so that level sets of the temperature are also level sets of the entropy, (which are also level sets of the potential). Introducing some convenient normalization constants $E_*$ and $T_*$, related by $E_* = k_B \, T_*$, this can be summarized by saying that the general solution of equation (\ref{E: key-1}) is:
\begin{equation}
T(\r) = {T_*\over f'(-\Phi(\r)/E_*)}; \qquad S= k_B \; f(-\Phi(\r)/E_*).
\label{E: entropy-force-1}
\end{equation}
Here $f(x)$ is an arbitrary monotonic function and $f'(x) = \d f/\d x$ is its derivative. (One can easily verify this solution is correct by using the chain rule, with the monotonicity of $f(x)$ being required to avoid a divide by zero error.) This very simple and very general constraint on the temperature and entropy of any thermodynamic system capable of mimicking an externally imposed conservative force is nevertheless very powerful --- and we shall soon see that this result is very problematic for Verlinde's proposal.

\subsection{$n$-body scenario} \label{SS: n-body}

When one considers $n$ bodies mutually  interacting via some conservative force,  the argument is very similar, with just enough difference to make an explicit exposition worthwhile.  The force on the $i^{th}$ particle is now
\begin{equation}
\F_i(\r_1,\dots,\r_n) = -\bnabla_i\Phi(\r_1,\dots,\r_n).
\end{equation}
If we \emph{assume} this can be mimicked by an entropic force then we must have
\begin{equation}
\F_i(\r_1,\dots,\r_n) = T(\r_1,\dots,\r_n) \; \bnabla_i S(\r_1,\dots,\r_n),
\end{equation}
implying
\begin{equation}
\bnabla_i\Phi(\r_1,\dots,\r_n) = - T(\r_1,\dots,\r_n) \; \bnabla_i S(\r_1,\dots,\r_n).
\label{E: key-n}
\end{equation}
Without any calculation, since $\forall i$ we have $\bnabla_i \Phi || \bnabla_i S$, this immediately implies that the level sets of the potential are also level sets of the entropy. (Thus implying that the entropy is some function of the potential.) But by taking the curl (with respect to the variable $\r_i$) of equation (\ref{E: key-n}) we also see that $\forall i$ we have $\bnabla_i T || \bnabla_i S$, so that level sets of the temperature are also level sets of the entropy, (which are also level sets of the potential). As in the 1-body scenario, introducing some convenient normalization constants $E_*$ and $T_*$, related by $E_* = k_B \, T_*$, this can be summarized by saying that the general solution of equation (\ref{E: key-n}) is:
\begin{equation}
T(\r_1,\dots,\r_n) = {T_*\over f'(-\Phi(\r_1,\dots,\r_n)/E_*)}; \qquad S(\r_1,\dots,\r_n)= k_B \; f(-\Phi(\r_1,\dots,\r_n)/E_*).
\label{E: entropy-force-2}
\end{equation}
Here $f(x)$ is again  an arbitrary monotonic function and $f'(x) = \d f/\d x$ is its derivative. (One can again easily verify this solution is correct by using the chain rule,  with the monotonicity of $f(x)$) being required to avoid a divide by zero error.) This very simple and very general constraint on the temperature and entropy of any thermodynamic system capable of mimicking the dynamics of $n$ bodies mutually interacting via a conservative force is nevertheless very powerful --- and we shall soon see that this result is very problematic for Verlinde's proposal.

\subsection{$n$-body Newtonian gravity} \label{SS: n-body-Newton}

In the specific case of Newtonian gravity we have
\begin{equation}
 \Phi(\r_1,\cdots,\r_n) = - {1\over2} \sum_{j\neq i} { G m_i m_j\over |\r_i - \r_j|},
\end{equation}
so that
\begin{equation}
T(\r_1,\dots,\r_n) = {T_*\over f'\left(\displaystyle{1\over2 E_*} \sum_{j\neq i} { G m_i m_j\over |\r_i - \r_j|}\right)},
\end{equation}
and
\begin{equation}
S(\r_1,\dots,\r_n)= k_B \; f\left( {1\over2 E_*} \sum_{j\neq i} { G m_i m_j\over |\r_i - \r_j|}\right).
\end{equation}
Specifically, this is enough to tell you that \emph{if} Newtonian gravity can be mimicked by a entropic force \emph{then}  (in view of the monotonicity of $f(x)$) the entropy must be high when the particles are close together.

Purely as an example, a very specific proposal is to take $f(x)=x$, in which case
\begin{equation}
T(\r_1,\dots,\r_n) = T_*;
\qquad 
S(\r_1,\dots,\r_n)= {k_B \over2 E_*} \sum_{j\neq i} { G m_i m_j\over |\r_i - \r_j|}.
\label{E: specific-newton}
\end{equation}
This is arguably the simplest possible entropic force model one could come up with for Newtonian gravity --- it certainly accurately reproduces the dynamics of Newtonian gravity, \emph{but it is very different in detail from Verlinde's proposal}. (One reason for possibly being interested in this specific proposal is that it is isothermal, and the known examples of entropic forces in condensed matter setting typically take place in an isothermal environment.)  

\subsection{$n$-body Coulomb force} \label{SS: n-body-coulomb}
 
In the specific case of the Coulomb force we have
\begin{equation}
 \Phi(\r_1,\cdots,\r_n) =  {1\over8\pi\epsilon_0} \sum_{j\neq i} { q_i q_j\over |\r_i - \r_j|},
\end{equation}
so that
\begin{equation}
T(\r_1,\dots,\r_n) = {T_*\over f'\left(\displaystyle-{1\over8\pi\epsilon_0\; E_*} \sum_{j\neq i} { q_i q_j\over |\r_i - \r_j|}\right)},
\end{equation}
and
\begin{equation}
S(\r_1,\dots,\r_n)= k_B \; f\left(-{1\over8\pi\epsilon_0\; E_*} \sum_{j\neq i} { q_i q_j\over |\r_i - \r_j|}\right).
\end{equation}
Specifically, this is enough to tell you that \emph{if} the Coulomb force can be mimicked by a entropic force \emph{then}  (in view of the monotonicity of $f(x)$, and the fact that the Coulomb potential is of indefinite \emph{sign}, and as long as we have $f(x_*)=0$ at some finite value $x_*$),  one must be prepared to deal with negative entropies and temperatures. Such a condition permits one, for instance, to choose a normalization such that  $S(\Phi=0)=0$. In the present context this normalization is very natural in the sense that it assigns zero entropy to infinitely dispersed systems. 

Now negative entropies and temperatures are outside the realm of classical thermodynamics, but are nevertheless well-established concepts in theoretical physics. Negative temperatures are common in statistical physics~\cite{Ramsey}, where they are a signal that one is encountering a ``population inversion'' (for example, in certain nuclear spin systems~\cite{Purcell-Pound}, in certain atomic gasses~\cite{Mosk},  or in laser physics~\cite{Kardar}). 
Negative entropies are less common, but ``negentropy'' is often interpreted in terms of ``information'' --- see for instance Shannon's information theory~\cite{Shannon}, and various attempts at reinterpreting thermodynamics in terms of information theory~\cite{Landauer, Brillouin}. (We shall subsequently see many more instances of negative entropies and negative temperatures when we explore Verlinde's specific approach later on in this article.)

Purely as an example, a very specific proposal is to take $f(x)=x$, in which case
\begin{equation}
T(\r_1,\dots,\r_n) = T_*;
\qquad 
S(\r_1,\dots,\r_n)= -{k_B\over8\pi\epsilon_0\; E_*} \sum_{j\neq i} { q_i q_j\over |\r_i - \r_j|}.
\label{E: specific-coulomb}
\end{equation}
This is arguably the simplest possible entropic force model one could come up with for the Coulomb force --- it certainly accurately reproduces the dynamics of the Coulomb force. (But note that this specific proposal is qualitatively different from Wang's proposal in~\cite{Wang}, and is at best orthogonal to Verlinde's suggestions in~\cite{Verlinde}.) 

\section{Problems with Verlinde's proposal}

The problems with Verlinde's proposal come from his specific suggestions for making the temperature depend on a non-relativistic variant of the Unruh effect~\cite{Unruh}, and making the entropy depend on the ``distance from a holographic screen''~\cite{Verlinde}. We have just seen that for conservative entropic forces we only have one free function $f(x)$ to play with, and that is simply not sufficient to satisfy all of Verlinde's requirements. 

\subsection{$1$-body external potential scenario} \label{SS: 1-body-problems}

For a single body interacting with an external potential, if the force is to be given an entropic interpretation then the temperature is necessarily some function of the potential, $\Phi$. But Verlinde wants the temperature to be interpretable in terms of a non-relativistic variant of the Unruh effect~\cite{Unruh}, so it must be some function of the norm of the acceleration $\a$, and so must be some function of the norm of the \emph{gradient} of the potential $|\bnabla\Phi|$.

But in general these requirements are mutually inconsistent: The level sets of the potential $\Phi$ are generically not the same as the level sets of the norm of the gradient of the potential $|\bnabla\Phi|$. The level sets coincide only under extremely stringent symmetry hypotheses (such spherical symmetry, cylindrical symmetry, or planar symmetry).  For a generic potential these level sets simply do not coincide, and if we want the force to have an entropic reinterpretation then the temperature of this hypothetical thermodynamic system cannot depend on $|\bnabla\Phi|$, and so cannot have a Unruh-effect interpretation. 

More explicitly, in the 1-body external potential scenario we must have $\bnabla S \propto \a$, and Verlinde's proposal amounts to making the identifications:
\begin{equation}
T = {\hbar \; |\a|\over 2\pi k_B c};   \qquad \bnabla S = {2\pi k_B m c\over\hbar} \; \bm{\hat a}.
\end{equation}
That is
\begin{equation}
T = {\hbar \; |\bnabla\Phi|\over 2\pi k_B m c};   \qquad \bnabla S = - {2\pi k_B m c\over\hbar} \; {\bnabla\Phi\over |\bnabla\Phi|}. 
\end{equation}
But this last equation, 
\begin{equation}
\bnabla S = \hbox{(const)} \; {\bnabla\Phi \over |\bnabla\Phi|},
\label{E: ill-posed}
\end{equation}
 is generically ill-posed. Given generic $\Phi$ as input, no such $S$ exists. It is only when the level sets of $\Phi$ coincide with the level sets of $|\bnabla\Phi|$ that this differential equation has solutions. (That is, the iso-potential surfaces have to coincide with the iso-acceleration surfaces. This is an extremely stringent constraint satisfied only in situations of very high symmetry, such as spherical, cylindrical, or planar symmetry.) Note that this is not an argument against entropic forces, nor even an argument against entropic reinterpretations of Newtonian gravity, it is instead an argument against Verlinde's specific proposals for $T$ and $\bnabla S$. 

Is there another assignment that would work? Yes, as we have seen, for arbitrary monotonic $f(x)$ the assignment
\begin{equation}
T = {T_*\over f'(-\Phi/E_*)}; \qquad S=k_B\; f(-\Phi/E_*),
\end{equation}
successfully does the job of reproducing the classical force $\F$, though in this proposal the temperature does not have any direct interpretation in terms of the Unruh effect.

\subsection{$2$-body scenario} \label{SS: 2-body-problems}

Somewhat different problems affect the $2$-body scenario. At the most basic level Verlinde's proposal would assign a different temperature to each particle
\begin{equation}
T_i = {\hbar \; |\a_i|\over 2\pi k_B c} = {\hbar \; |\bnabla_i\Phi|\over 2\pi k_B m_i c},   \qquad\qquad i\in\{1,2\},
\label{E: Unruh-2}
\end{equation}
whereas the standard notion of entropic force really only has room for a single temperature to be assigned to the whole thermodynamic system. 

If we put this aside for now, and concentrate on the entropy, Verlinde's key axiom is that a particle near a ``holographic screen'' in some sense contributes an entropy~\cite{Verlinde}
\begin{equation}
\Delta S = 2\pi k_B \; {m c\Delta x\over\hbar}.
\label{E: entropy-axiom}
\end{equation}
Verlinde takes the entropy to increase as the particle moves \emph{towards} the ``holographic screen''. 
Let us call $S_0$ the entropy of the ``holographic screen''   when the particle is located on the screen itself, and $\ell$ the geodesic distance to the screen. Then at least for small $\ell$ we can formalize this as
\begin{equation}
S = S_0 -  2\pi k_B \; {m c\ell\over\hbar},
\label{E: entropy-axiom2}
\end{equation}
or even
\begin{equation}
\bnabla S = - {2\pi k_B m c\over \hbar}\;  \bm{\hat n}.
\label{E: entropy-axiom3}
\end{equation}
Here $\bm{\hat n}$ is the ``outward normal to the holographic screen''.  The minus sign is important. 

For two particles we have two masses $m_i$.  \emph{As long as we are dealing with a central force}, in a 2-body system it is appropriate to choose two spherical ``holographic screens'', one  around each particle individually, thereby defining two normal vectors $\bm{\hat n}_i$.
This strongly suggests that we need two entropies
\begin{equation}
\bnabla_i S_i = - {2\pi k_B m_i c\over \hbar}\;   \bm{\hat n}_i, \qquad i\in\{1,2\}, \qquad \hbox{(no sum on $i$)}.
\label{E: whatever}
\end{equation}
But then, as long as we are dealing with a central force, in a 2-body system
\begin{equation}
\bm{\hat n}_i \; || \; (\r_i-\r_{\mathrm{not}(i)}), \qquad i\in\{1,2\}.
\end{equation}
Because of the very high symmetry, in the 2-body situation we can integrate these two equations to obtain (up to irreverent constants of integration) the entropies:
\begin{equation}
S_i = - {2\pi k_B m_i c\over \hbar}\;   |\r_i-\r_{\mathrm{not}(i)}|, \qquad \qquad i\in\{1,2\}.
\label{E: whatever2}
\end{equation}
Note that these entropies are negative. (Even if we had used the arbitrary constants of integration to make the entropy positive at zero separation, one would nevertheless be driven to negative entropy at large separation. So in Verlinde's ``holographic screen'' framework for entropy, there is no real loss of either generality or ``physical reasonableness'' in choosing to normalize these entropies to zero at zero separation.)
To reproduce the 2-body force law we must now take
\begin{equation}
\F_i = T_i \; \bnabla_i S_i \qquad \qquad i\in\{1,2\}, \qquad \hbox{(no sum on $i$)}.
\label{E: whatever3}
\end{equation}
Unwrapping this expression for the force we see
\begin{equation}
\F_i =  - |\bnabla_i \Phi| \; \n_i.
\end{equation}
But there are various ways in which this proposal still does not quite work.
\begin{itemize}
\item For  Newtonian gravity this version of the proposal at least generates an attractive 2-body force, at the cost of negative entropies  $S_i$. (Though the temperatures $T_i$ are at least positive.) 
\item For a Coulomb 2-body situation one needs an  additional \emph{ad hoc} ``fix''. To keep track of attraction versus repulsion one needs to take the sign of the individual charges $q_i$ outside the $|\a_i|$ of equation (\ref{E: Unruh-2}), so that $|\a_i|\to -sign(q_1 q_2) |\a_i|$, and $T_i \to -sign(q_1 q_2) \; |T_i|$.  Then
\begin{equation}
\F_i =  sign(q_1 q_2) |\bnabla_i \Phi| \; \n_i.
\end{equation}
Note that for charges of the same sign (repulsive electric forces) one now needs negative temperatures.  We shall subsequently (see equation (\ref{E: n-body-fix-T})) find a better way of dealing with this ``fix'' to make it less \emph{ad hoc}. 
\item More disturbingly, the very notion of needing to use two temperatures, and two entropies,  to reproduce 2-body Newtonian gravity is rather orthogonal to standard notions of entropic force. 
\end{itemize}
Why were these issues unrecognized Verlinde's article? Because the explicit calculations carried out there did not look at the 2-body scenario, and dealt exclusively with the test particle limit. (And even more restrictively with the test particle limit in situations of extremely high symmetry.)

\subsection{$(n\geq3)$-body scenario} \label{SS: n-body-problems}

Related but even more acute problems affect the $n$-body scenario. For $n\geq 3$ one has to deal \emph{both} with multiple temperatures, 
\begin{equation}
T_i = {\hbar \; |\a_i|\over 2\pi k_B c} = {\hbar \; |\bnabla_i\Phi|\over 2\pi k_B m_i c},   \qquad\qquad i\in\{1,\dots,n\},
\label{E: Unruh-n}
\end{equation}
\emph{and} with ill-posed differential equations determining the entropies $S_i$.
At various points of his article~\cite{Verlinde}, Verlinde rather strongly suggests that his ``holographic screens'' be located on equipotential surfaces, in which case the normal appearing in equation (\ref{E: entropy-axiom3}) is $\bm{\hat n} = \bnabla\Phi/|\bnabla\Phi|$. But then we are back to the equation
\begin{equation}
\bnabla S = - {2\pi k_B m c\over \hbar}\;  {\bnabla\Phi \over |\bnabla\Phi|},
\end{equation}
which we had previously seen is generically ill-posed. (That is, ill-posed except in situations of extremely high symmetry.) In fact, one should write down one such equation for each individual particle,
\begin{equation}
\bnabla_i S_i = - {2\pi k_B m_i c\over \hbar}\;  {\bnabla_i\Phi \over |\bnabla_i\Phi|}, \qquad i\in\{1,\cdots,n\}, \qquad \hbox{(no sum on $i$)}.
\label{E: whatever4}
\end{equation}
But for $n\geq3$ bodies the potential $\Phi(\r_1,\dots,\r_n)$ generically has no symmetries,  so these are ill-posed equations that generically have no solutions. 

We have gone through these problematic issues in some detail because the problems raised now give us some hints on how to proceed.  I again emphasize that I am not particularly worried about entropic forces \emph{per se}, it is instead the combination of entropic forces with the Unruh effect and ``holographic screens'' that leads to problems. 

\section{Thermodynamic forces}

Suppose we have a complicated thermodynamic system that is described by a large number of intensive variables $x_a$ and a correspondingly large number of extensive variables $X_a$. Then one can write down an expression for a ``thermodynamic force''
\begin{equation}
\F = \sum_a x_a \; \bnabla X_a,
\end{equation}
which now has a more general structure than that normally assigned to an ``entropic force''.  This sort of decomposition is much more promising when it comes to a coherent implementation of Verlinde's ideas within the thermodynamic force scenario.

\subsection{$1$-body external potential scenario} \label{SS: 1-body-fix}

We had previously seen that the differential equation (\ref{E: ill-posed}) was ill-posed unless the potential was of very high symmetry. So let us assume that the potential decomposes into a linear sum of such highly symmetric potentials
\begin{equation}
\Phi(\r) = \sum_a \Phi_a(\r).
\end{equation}
Let the individual $\Phi_a(\r)$ be either spherically symmetric, cylindrically symmetric, or plane symmetric. Let $\ell_a$ denote the geodesic distance to the centre of the spherically symmetric potentials, the geodesic distance to the axis of the cylindrically symmetric potentials, and the (signed) geodesic distance to some convenient plane of symmetry for the plane symmetric potentials. Then by construction for each individual potential we have $\Phi_a(r)=\Phi_a(\ell_a)$.  For each individual potential $\Phi_a$ we can now integrate the differential equations
\begin{equation}
\bnabla S_a = - {2\pi k_B m c\over \hbar}\; \n_a =- {2\pi k_B m c\over \hbar}\;  \bnabla\ell_a,
\end{equation}
to yield (up to arbitrary irrelevant constants of integration):
\begin{equation}
S_a = - {2\pi k_B m c\over \hbar}\;  \ell_a.
\label{E: 1-body-fix-S}
\end{equation}
If we now define
\begin{equation}
T_a= - {\hbar \over 2\pi k_B c} \; ( \a_a \cdot \n_a) = {\hbar \; \over 2\pi k_B m c} \; {\partial\Phi_a\over \partial {\ell_a}},  
\qquad \hbox{(no sum on $a$)}, 
\label{E: Unruh-1-body}
\end{equation}
then, as required
\begin{equation}
\F = \sum_a T_a \; \bnabla S_a = - \sum_a {\partial\Phi_a\over \partial {\ell_a}} \; \bnabla\ell_a = -\bnabla\left({\sum\nolimits_a \Phi_a}\right) = - \bnabla \Phi.
\end{equation}
Several comments are in order:
\begin{itemize}
\item 
This is indeed a ``thermodynamic'' interpretation of the force, but now with an unboundedly large number of ``temperatures'' $T_a$, and ``entropies'' $S_a$. 
This lies well outside the usual notion of entropic force, but seems to be the minimum requirement if one wants to explain/reinterpret externally applied conservative forces as ``thermodynamic forces'' while simultaneously having temperatures that are ``Unruh-like'' and an entropy that is ``holographic'' in Verlinde's sense. 
\item
Note the use of $\a_a \cdot \n_a$ rather than $|\a_a|$, and $ {\partial\Phi_a/ \partial {\ell_a}}$ rather than $\bnabla\Phi_a$. This is done to automatically take care of the signs for attractive and repulsive potentials, so the formalism works equally well for gravity and electromagnetism --- the formalism can now even handle potentials such as the Lennard--Jones potential where the force can change sign as a function of distance.
\item
Note that for attractive forces the Unruh-like temperature $T_a$ is positive, while for repulsive forces it is negative. This explains (and makes systematic) the othewise \emph{ad hoc} fix we encountered when considering 2-body Coulomb forces. 
\item
Note that the physical 3-acceleration satisfies
\begin{equation}
\a = \sum_a \a_a.
\end{equation}
So based loosely on the Unruh effect one might define a ``total temperature''
\begin{equation}
T = \left|  \sum\nolimits_a T_a \; \n_a  \right|  \leq  \sum\nolimits_a  \left| T_a\right|.
\end{equation}
The utility of such a definition is uncertain.
\item
One might also try to define a ``total entropy''
\begin{equation}
S = \sum_a S_a =  - {2\pi k_B mc \over \hbar} \; \sum_a \ell_a. 
\end{equation}
The utility of such a definition is uncertain. In particular $ \sum_a \ell_a$ does not seem to have a clear physical interpretation/justification.
\end{itemize}

\subsection{$n$-body scenario} \label{SS: n-body-fix}

Having seen what can be done for a external potential once we adopt multiple ``temperatures'' and ``entropies'', and having seen what goes wrong with the most naive implementations of the 2-body and $n$-body potentials, we can now see our way to a general ``thermodynamic'' ansatz that at least reproduces the classical force we are seeking to emulate. 
Consider any $n$-body potential that is a linear sum of 2-body central potentials:
\begin{equation}
\Phi(\r_1,\dots,\r_n)= {1\over2}\sum_{i,j}^{i\neq j} \Phi_{ij}(\r_i-\r_j).
\end{equation}
For each \emph{ordered pair} of particles, based on the 2-body results of the previous section,  postulate
\begin{equation}
S_{i:j} = -{2\pi k_B m_i c\;  |\r_i - \r_j| \over \hbar} =  -{2\pi k_B m_i c\; \ell_{ij} \over \hbar},    \qquad\qquad   i,j \in\{1,\dots,n\}. 
\label{E: n-body-fix-S}
\end{equation}
Note the absence of any interchange symmetry; this is the ``entropy'' of particle $i$ due to the presence of particle $j$. (I shall use the subscript ``$i\!:\!j$'' to emphasise the lack of symmetry under particle interchange, and use ``$ij$'' whenever the quantity is symmetric under particle interchange.) 
Based \emph{very loosely} on the Unruh effect, one can argue that there is also a ``temperature'' of particle  $i$ due to the presence of particle $j$:
\begin{equation}
T_{i:j} = - {\hbar \over2\pi k_B c} (\a_{i:j}\cdot \n_{i:j}) = {\hbar \; \over 2\pi k_B m_i c} \; {\partial\Phi_{ij}\over \partial {\ell_{ij}}},    \qquad\qquad   i,j \in\{1,\dots,n\}. 
\label{E: n-body-fix-T}
\end{equation}
Again note the absence of any interchange symmetry.
Effectively one is temporarily pretending all other particles are absent, calculating what the acceleration of particle $i$ would be if particle $j$ were the only other particle in the universe, and using that acceleration $\a_{i:j}$ in Unruh's formula~\cite{Unruh} to define the temperature $T_{i:j}$. Because of the assumption that we are dealing with a central force, we have  $(\a_{i:j}\cdot \n_{i:j}) = \mp |\a_{i:j}|$, with $-$ for attraction and $+$ for repulsion.
The force on particle $i$ is
\begin{eqnarray}
\F_i &=& \sum_j^{j\neq i} T_{i:j} \; \bnabla_i S_{i:j} 
= - \sum_j^{j\neq i} {\partial\Phi_{ij}\over \partial {\ell_{ij}}} \; \n_{i:j} 
= - \sum_j^{j\neq i} {\partial\Phi_{ij}\over \partial {\ell_{ij}}} \; \bnabla_i\ell_{ij} 
\nonumber\\
&=& 
 - \bnabla_i \left(\sum\nolimits_j^{j\neq i} \Phi_{ij} \right) 
 =- \bnabla_i \left( {1\over2} \sum\nolimits_{i,j}^{j\neq i} \Phi_{ij} \right) 
 = -\bnabla_i \Phi(\r_1,\dots,\r_n).
\end{eqnarray}
Note that this construction works for any $n$-body potential that is a linear sum of 2-body central potentials. Both attractive and repulsive forces are automatically dealt with by phrasing the ``temperatures'' in terms of $\partial_\ell \Phi$, (rather than $|\partial_\ell\Phi|$). 

This construction at least reproduces the classical force law we are attempting to emulate using thermodynamic means. But,  to paraphrase Alice, consider the number of impossible things one  has to believe in before breakfast:
\begin{itemize}
\item 
You need a whole collection of $n(n-1)$ ``temperatures'' $T_{i:j}$, one for each \emph{ordered pair} of particles,  which do not add in any sensible way.  
Note that the 3-accelerations of the individual particles now satisfy
\begin{equation}
\a_i = \sum_j^{j\neq i} \a_{i:j}.
\end{equation}
So based rather loosely on the Unruh effect one might guess that each individual particle can be assigned a ``temperature'':
\begin{equation}
T_i = \left|  \sum\nolimits_j^{j\neq i} T_{i:j} \; \n_{i:j} \right| \leq  \sum\nolimits_j^{j\neq i} |T_{i:j}|.
\end{equation}
But there seems to be no sensible way of defining an overall ``temperature'' for the entire $n$-body system. 
\item 
You also need a whole collection of $n(n-1)$ entropies $S_{i:j}$, one for each \emph{ordered pair} of particles.  
For the total entropy $S$, if we boldly assert
\begin{equation}
S = \sum_{i,j}^{j\neq i} S_{i:j} = - \sum_{i,j}^{j\neq i} {2\pi k_B m_i c\;  |\r_i - \r_j| \over \hbar},
\end{equation}
then defining $R=\max_{ij}\{\;  |\r_i - \r_j| \}$, we see that with this definition we have
\begin{equation}
|S| \leq \sum_{i,j}^{j\neq i} {2\pi k_B m_i c\;  R \over \hbar}  = {2 \pi k_B M c R\over\hbar}.
\end{equation}
So \emph{up to a sign}, this prescription at least provides a Newtonian version of the Bekenstein bound~\cite{Bekenstein}. Note that this bound would apply to any central force that is given a thermodynamic interpretation in terms of ``holographic screens'' using the pairwise entropies $S_{i:j}$ above. This bound is not specific to gravity, either Newtonian or general relativistic.
Whether or not this observation has any deeper significance is unclear. 
\end{itemize}

\subsection{$n$-body Newton and Coulomb forces}

In this section we have seen that it is possible to find an interpretation of Verlinde's ideas that simultaneously is ``thermodynamic'', respects the Unruh-like interpretation of temperature and is compatible with Verlinde's  ``holographic screens'', and correctly reproduces the original classical force that one is attempting to emulate. \emph{But the price paid for this is very high}.  One has to introduce \emph{multiple} ``temperatures'' and ``entropies'',  one for each \emph{ordered pair} of particles, whose physical interpretation is far from clear.  In the specific case of Newtonian gravity we should take
\begin{equation}
S_{i:j}  =  -{2\pi k_B m_i c\; \ell_{ij} \over \hbar},    \qquad\qquad   i,j \in\{1,\dots,n\},
\end{equation}
and 
\begin{equation}
T_{i:j}  = {\hbar \; \over 2\pi k_B c} \; {G m_j\over \ell_{ij}^2},  \qquad\qquad   i,j \in\{1,\dots,n\}.
\label{E: n-body-fix-newton}
\end{equation}
In the case of the Coulomb force the ``entropies'' remain the same, but the ``temperatures'' are modified to be
\begin{equation}
T_{i:j}  = - {\hbar \; \over 2\pi k_B m_i c} \; {q_i q_j\over 4\pi\epsilon_0\; \ell_{ij}^2},  \qquad\qquad   i,j \in\{1,\dots,n\}.
\label{E: n-body-fix-coulomb}
\end{equation}
Note that ``entropies'' are negative, while for the Coulomb force the ``temperatures'' can be either positive or negative depending on whether the two particles in the pair are equally or oppositely charged.  Though somewhat complicated, this particular assignment of multiple ``temperatures'' and ``entropies'' seems to be the \emph{minimum requirement} to make something like Verlinde's suggestions work.

\section{Relative accelerations and reduced masses}

There is a minor variant of the formalism that is a little more symmetric, but at the cost of moving somewhat further from any usual interpretation of the Unruh effect, and that is to work with pairwise \emph{relative accelerations} and pairwise \emph{reduced masses}. That is, consider the antisymmetric quantity
\begin{equation}
\Delta \a_{ij} = \a_{i:j} - \a_{j:i} 
\end{equation}
and the symmetric quantity
\begin{equation}
\mu_{ij} =  {m_i \, m_j\over m_i+m_j}.
\end{equation}
Then choose
\begin{equation}
T_{ij} = - {\hbar \over2\pi k_B c} (\Delta \a_{ij}\cdot \n_{i:j}) = {\hbar \; \over 2\pi k_B \mu_{ij} c} \; {\partial\Phi_{ij}\over \partial {\ell_{ij}}},    \qquad\qquad   i,j \in\{1,\dots,n\},
\label{E: T-reduced}
\end{equation}
and
\begin{equation}
S_{ij} = -{2\pi k_B \mu_{ij} c\;  |\r_i - \r_j| \over \hbar} =  -{2\pi k_B \mu_{ij} c\; \ell_{ij} \over \hbar},    \qquad\qquad   i,j \in\{1,\dots,n\}. 
\label{E: S-reduced}
\end{equation}
These temperatures and entropies are now symmetric under particle interchange, so one needs only ${1\over2}n(n-1)$ heat baths. With this assignment one again thermodynamically reproduces the force law one is aiming for, now in a more symmetric fashion, and reducing the number of required  heat baths by a factor of 2. But the price again is very high --- in assigning the temperature $T_{ij}$ one not only has to ignore all the other particles in the universe, but one has to decompose the motion of the $ij$ pair into centre of mass and relative motions, and then to assign the temperature $T_{ij}$ to a fictitious ``particle'' mimicking the relative acceleration. By this stage the relationship of this $T_{ij}$ to Unruh's derivation of acceleration radiation~\cite{Unruh} is becoming \emph{severely} strained. The relationship of this $S_{ij}$ to Verlinde's proposals regarding holographic screens is also becoming rather strained. Furthermore, because the quantity $\sum_{i\neq j} \mu_{ij} \ell_{ij}$ has no particularly pleasant features, for the total entropy one severely degrades the tentative connection to the Bekenstein bound. 

If we nevertheless proceed along these lines, then in the specific case of Newtonian gravity we should take
\begin{equation}
S_{ij}  =  -{2\pi k_B \mu_{ij} c\; \ell_{ij} \over \hbar},    \qquad\qquad   i,j \in\{1,\dots,n\},
\end{equation}
and 
\begin{equation}
T_{ij}  = {\hbar \; \over 2\pi k_B c} \; {G (m_i+m_j)\over \ell_{ij}^2},  \qquad\qquad   i,j \in\{1,\dots,n\}.
\label{E: n-body-fix-newton2}
\end{equation}
In the case of the Coulomb force the ``entropies'' remain the same, but the ``temperatures'' are modified to be
\begin{equation}
T_{ij}  = - {\hbar \; \over 2\pi k_B \mu_{ij} c} \; {q_i q_j\over 4\pi\epsilon_0\; \ell_{ij}^2},  \qquad\qquad   i,j \in\{1,\dots,n\}.
\label{E: n-body-fix-coulomb2}
\end{equation}

\section{Further extensions of the formalism}

If the only thing one is interested in doing is to somehow mimic a classical force field by thermodynamic means, then one has considerable flexibility. For instance, the prescriptions of equations (\ref{E: entropy-force-1}) and (\ref{E: entropy-force-2}) provide an entropic force interpretation of any conservative force in terms of a single heat bath, but those prescriptions do not have any Unruh-like interpretation for the temperature, nor is there any natural way to introduce ``holographic screens'' into that formalism.  In contrast, if one wishes to implement some version of Verlinde's ideas, one is forced into quite complicated constructions using multiple intensive and extensive thermodynamic parameters; multiple ``temperatures'' and ``entropies''. See equations~(\ref{E: 1-body-fix-S})--(\ref{E: Unruh-1-body}),   and (\ref{E: n-body-fix-S})--(\ref{E: n-body-fix-T}),  and more specifically (\ref{E: n-body-fix-newton})--(\ref{E: n-body-fix-coulomb}).

If one is willing to relax Verlinde's conditions then there is tremendous flexibility in the thermodynamic force scenario. For instance,  pick some arbitrary collection of dimensionless monotonic functions $h_{i:j}(x)$, and a convenient length scale $L_*$,  and postulate
\begin{equation}
S_{i:j} = -{2\pi k_B m_i c\;  L_* \; h_{i:j} ( \; |\r_i - \r_j| /L_*\; ) \over \hbar}, \qquad\qquad   i,j \in\{1,\dots,n\}.
\end{equation}
Note the absence of symmetry; this is to be interpreted as the ``entropy'' of particle $i$ due to the presence of particle $j$.
\\
For the Newton force postulate the ``temperatures''
\begin{equation}
T_{i:j} =  {\hbar  \over2\pi c k_B } \; { G m_j\over |\r_i - \r_j|^2  \; h'_{i:j}( \; |\r_i - \r_j| /L_*\; )}, \qquad\qquad   i,j \in\{1,\dots,n\}.
\end{equation}
For the Coulomb force postulate
\begin{equation}
T_{i:j} =  {\hbar  \over2\pi m_i c k_B} \; {q_i q_j \over4\pi\epsilon_0\; |\r_i - \r_j|^2  \; h'_{i:j}( \; |\r_i - \r_j| /L_*\; )}, \qquad\qquad   i,j \in\{1,\dots,n\}.
\end{equation}
For a general central potential postulate
\begin{equation}
T_{i:j} =  {\hbar  \over2\pi m_i c k_B} \; {  \Phi_{ij}'(\; |\r_i - \r_j| \; ) \over h'_{i:j}( \; |\r_i - \r_j| /L_*\; )}, \qquad\qquad   i,j \in\{1,\dots,n\}.
\end{equation}
Again note the absence of any symmetry.
Then
\begin{eqnarray}
\F_i = \sum_j^{j\neq i} T_{i:j} \; \bnabla_i S_{i:j} = \sum_j^{j\neq i}    \Phi_{ij}'(\; |\r_i - \r_j| \; )  \bnabla_i |\r_i - \r_j|
= \sum_j^{j\neq i}    \bnabla_i  \Phi_{ij}(\; |\r_i - \r_j| \; ).
\end{eqnarray}
Note that (after application of the chain rule) the functions $h_{i:j}'(\cdot)$ and the constant $L_*$ have cancelled.  One then has
\begin{eqnarray}
\F_i  =   \bnabla_i  \left(\sum\nolimits_j^{j\neq i}   \Phi_{ij}(\; |\r_i - \r_j| \; ) \right) 
= \bnabla_i  \left( {1\over2} \sum\nolimits_{i,j}^{j\neq i}   \Phi_{ij}(\; |\r_i - \r_j| \; ) \right), 
\end{eqnarray}
whence finally
\begin{eqnarray}
\F_i  =\bnabla_i\Phi(\r_1,\cdots,\r_n).
\end{eqnarray}
This now  reproduces the classical force law we are trying to emulate by thermodynamic methods.  

Setting $h(x)=x$ reproduces the Verlinde-like proposal of equations (\ref{E: n-body-fix-S})--(\ref{E: n-body-fix-T}),  and more specifically of equations (\ref{E: n-body-fix-newton}) and  (\ref{E: n-body-fix-coulomb}) above. On the other hand, if we choose
\begin{equation}
h_{i:j}(\ell_{ij}/L_*) = {\Phi_{ij}(\ell_{ij}) \over m_i c^2},
\end{equation}
then
\begin{equation}
T_{i:j} = {\hbar c\over2\pi k_B L_*} = T_*,
\end{equation}
and 
\begin{equation}
S_{i:j} = - {2\pi k_B L_* \Phi_{ij} \over \hbar c} = - k_B \; {\Phi_{ij}\over E_*}.
\end{equation}
We now have a single heat bath at a fixed temperature, $T_*$ and can define a total entropy
\begin{equation}
S = {1\over2} \sum_{i,j}^{i\neq j} S_{i:j} = - k_B \; {1\over2E_*} \sum_{i,j}^{i\neq j} \Phi_{ij} =  - k_B \; {\Phi\over E_*}. 
\end{equation}
This has now reproduced the specific models of equations (\ref{E: specific-newton}) and (\ref{E: specific-coulomb}). 
By using \emph{both} of the arbitrary monotonic functions $f(\cdot)$ and $h(\cdot)$ this could be generalized even further, but there seems little need (or utility) for doing so. 

\section{Brownian motion, heat baths, and decoherence}

So far, we have addressed the \emph{formal} question of how to represent conservative forces in an entropic manner. There are also serious \emph{physical} questions to be answered in any entropic force scenario. A particularly cogent criticism of the physical reality of the entropic force scenario is based on issues related to quantum mechanical collapse of the wavefunction~\cite{Kobakhidze, Kobakhidze:2011}. 
Stripped to its essentials, and provided we postulate the physical reality of entropic gravity, the issue is this:
\begin{itemize}
\item The fact that we do not see any Brownian noise superimposed on the motion of falling bodies (even individual elementary particles) indicates that the mean free time between interactions with the heat bath is very small, smaller than the temporal resolution of our experiments.
\item This indicates that one has a strong coupling to the heat bath.
\item But a strong coupling to the environment will cause decoherence --- ``collapse of the wavefunction'' --- on a timescale similar to the mean free time between interactions with the heat bath.
\item But quantum effects in external gravitational fields have been experimentally observed, without any sign of appreciable decoherence.  (See the technical discussion in~\cite{Kobakhidze, Kobakhidze:2011}.)
\end{itemize}
This suggests significant problems for the notion of gravity as an entropic force, at least insofar as we take the heat bath (or multiple heat baths) to be real and physical, not just convenient descriptive fictions.

\section{Discussion} \label{S: discussion}

In this article I have not attempted to justify reinterpreting Newtonian gravity as an entropic force, instead I have asked the question: ``\emph{If we assume Newtonian gravity is an entropic force, what does this tell us about the relevant thermodynamic system?}''  What can we say about the relevant temperature and entropy functions?  What constraints do they satisfy? The answers we have obtained are mixed:
\begin{itemize}
\item 
If we want to use a single heat bath, then any conservative force can be recast into entropic force form --- but the resulting model is at best orthogonal to Verlinde's proposal.
\item 
If we wish to retain key parts of Verlinde's proposal (an Unruh-like temperature, and entropy related to ``holographic screens''), then one is unavoidably forced into a more general ``thermodynamic force'' scenario with multiple intensive and extensive thermodynamic variables. (Multiple ``temperatures'' and ``entropies''.) 
The relevant ``entropies'' are negative, while the ``temperatures'' are positive for attractive forces and negative for repulsive forces. These features are certainly odd, and certainly not what might naively be expected. 
\end{itemize}
There is no reasonable doubt concerning the physical reality of entropic forces, and no reasonable doubt that classical (and semi-classical) general relativity is closely related to thermodynamics~\cite{Bekenstein:1973, Mechanix, Hawking:1974, Hawking:1975}. Based on the work of Jacobson~\cite{Jacobson, Jacobson:1999, Jacobson-Parentani, Eling, Eling:2008, Chirco}, Padmanabhan~\cite{Padmanabhan, Paranjape,  Mukhopadhyay, Kothawala, Kothawala:2009, Kolekar}, and others, there are also good reasons to suspect a thermodynamic interpretation of the fully relativistic Einstein equations might be possible.    Whether the specific proposals of Verlinde~\cite{Verlinde} are anywhere near as fundamental is yet to be seen --- the rather baroque construction needed to accurately reproduce $n$-body Newtonian gravity in a Verlinde-like setting certainly gives one pause. 

\section*{Acknowledgments}

I wish to thank Eolo Di Casola, Steve Carlip, Ted Jacobson, Fay Dowker, Raf Guedens, Stefano Liberati, Bei-Lok Hu, Thanu Padmanabhan, Silke Weinfurtner, and other participants of GTC2011 (SISSA, Trieste) for their comments. I also wish to thank Kane O'Donnell for his comments. This research was supported by the Marsden Fund administrated by the Royal Society of New Zealand. 

\clearpage



\begin{thebibliography}{99}

\bibitem{Jacobson}
  T.~Jacobson,
  ``Thermodynamics of space-time: The Einstein equation of state'',
  Phys.\ Rev.\ Lett.\  {\bf 75 } (1995)  1260-1263.
  [gr-qc/9504004].
  
  \bibitem{Jacobson:1999}
  T.~Jacobson,
  ``On the nature of black hole entropy'',
  [gr-qc/9908031].
  
  
 \bibitem{Jacobson-Parentani}
  T.~Jacobson, R.~Parentani,
  ``Horizon entropy'',
  Found.\ Phys.\  {\bf 33 } (2003)  323-348.
  [gr-qc/0302099].

  
\bibitem{Eling}
  C.~Eling, R.~Guedens, T.~Jacobson,
  ``Non-equilibrium thermodynamics of spacetime'',
  Phys.\ Rev.\ Lett.\  {\bf 96 } (2006)  121301.
  [arXiv:gr-qc/0602001 [gr-qc]].  
  
  \bibitem{Eling:2008}
  C.~Eling,
  ``Hydrodynamics of spacetime and vacuum viscosity'',
  JHEP {\bf 0811} (2008) 048.
  [arXiv:0806.3165 [hep-th]].
  
  \bibitem{Chirco}
  G.~Chirco, S.~Liberati,
  ``Non-equilibrium thermodynamics of spacetime: The role of gravitational dissipation'',
  Phys.\ Rev.\  {\bf D81 } (2010)  024016.
  [arXiv:0909.4194 [gr-qc]].
  
  
  
  \bibitem{Padmanabhan}
  T.~Padmanabhan,
  ``Is gravity an intrinsically quantum phenomenon? Dynamics of gravity from the entropy of space-time and the principle of equivalence'',
  Mod.\ Phys.\ Lett.\  {\bf A17 } (2002)  1147-1158.
  [hep-th/0205278].
  \\
  T.~Padmanabhan,
  ``Gravity from space-time thermodynamics'',
  Astrophys.\ Space Sci.\  {\bf 285 } (2003)  407.
  [gr-qc/0209088].
  \\
  T.~Padmanabhan,
  ``Gravitational entropy of static space-times and microscopic density of states'',
  Class.\ Quant.\ Grav.\  {\bf 21 } (2004)  4485-4494.
  [arXiv:gr-qc/0308070 [gr-qc]]. 
  \\
    T.~Padmanabhan,
  ``Gravity and the thermodynamics of horizons'',
  Phys.\ Rept.\  {\bf 406 } (2005)  49-125.
  [gr-qc/0311036].
  \\
  T.~Padmanabhan,
  ``Holographic gravity and the surface term in the Einstein-Hilbert action'',
  Braz.\ J.\ Phys.\  {\bf 35 } (2005)  362-372.
  [gr-qc/0412068].
  \\
  T.~Padmanabhan,
  ``A New perspective on gravity and the dynamics of spacetime'',
  Int.\ J.\ Mod.\ Phys.\  {\bf D14 } (2005)  2263-2270.
  [gr-qc/0510015].
  \\
  T.~Padmanabhan,
  ``Gravity: A new holographic perspective'',
  Int.\ J.\ Mod.\ Phys.\  {\bf D15 } (2006)  1659-1676.
  [gr-qc/0606061].
  \\
  T.~Padmanabhan,
  ``Gravity as an emergent phenomenon: A conceptual description'',
  AIP Conf.\ Proc.\  {\bf 939 } (2007)  114-123.
  [arXiv:0706.1654 [gr-qc]].
  \\
  T.~Padmanabhan,
  ``Entropy density of spacetime and thermodynamic interpretation of field equations of gravity in any diffeomorphism invariant theory'',
  [arXiv:0903.1254 [hep-th]].
  \\
  T.~Padmanabhan,
  ``Thermodynamical aspects of gravity: New insights'',
  Rept.\ Prog.\ Phys.\  {\bf 73 } (2010)  046901.
  [arXiv:0911.5004 [gr-qc]].
 \\
   T.~Padmanabhan,
  ``Surface density of spacetime degrees of freedom from equipartition law in theories of gravity'',
  Phys.\ Rev.\  {\bf D81 } (2010)  124040.
  [arXiv:1003.5665 [gr-qc]].
  \\
  T.~Padmanabhan,
  ``Lessons from classical gravity about the quantum structure of spacetime'',
  J.\ Phys.\ Conf.\ Ser.\  {\bf 306 } (2011)  012001.
  [arXiv:1012.4476 [gr-qc]].
  

  
  \bibitem{Paranjape}
  A.~Paranjape, S.~Sarkar, T.~Padmanabhan,
  ``Thermodynamic route to field equations in Lancos-Lovelock gravity'',
  Phys.\ Rev.\  {\bf D74 } (2006)  104015.
  [hep-th/0607240].
  
  \bibitem{Mukhopadhyay}
  A.~Mukhopadhyay, T.~Padmanabhan,
  ``Holography of gravitational action functionals'',
  Phys.\ Rev.\  {\bf D74 } (2006)  124023.
  [hep-th/0608120].
  
  \bibitem{Kothawala}
  D.~Kothawala, S.~Sarkar, T.~Padmanabhan,
  ``Einstein's equations as a thermodynamic identity: The cases of stationary axisymmetric horizons and evolving spherically symmetric horizons'',
  Phys.\ Lett.\  {\bf B652 } (2007)  338-342.
  [gr-qc/0701002].
  
  \bibitem{Kothawala:2009}
  D.~Kothawala, T.~Padmanabhan,
  ``Thermodynamic structure of Lanczos-Lovelock field equations from near-horizon symmetries'',
  Phys.\ Rev.\  {\bf D79 } (2009)  104020.
  [arXiv:0904.0215 [gr-qc]].
  
  \bibitem{Kolekar}
  S.~Kolekar, T.~Padmanabhan,
  ``Holography in Action'',
  Phys.\ Rev.\  {\bf D82 } (2010)  024036.
  [arXiv:1005.0619 [gr-qc]].
  

  \bibitem{Hu:2002}
  B.~L.~Hu,
  ``A Kinetic theory approach to quantum gravity'',
  Int.\ J.\ Theor.\ Phys.\  {\bf 41 } (2002)  2091-2119.
  [gr-qc/0204069].
 
 \bibitem{Hu:1999}
  B.~L.~Hu,
  ``Stochastic gravity'',
  Int.\ J.\ Theor.\ Phys.\  {\bf 38 } (1999)  2987-3037.
  [gr-qc/9902064]. 
  
  \bibitem{Hu-Verdaguer}
  B.~L.~Hu, E.~Verdaguer,
  ``Stochastic gravity: Theory and applications'',
  Living Rev.\ Rel.\  {\bf 11 } (2008)  3.
  [arXiv:0802.0658 [gr-qc]].
  
  
   
   \bibitem{Sakharov}
  A.~D.~Sakharov,
  ``Vacuum quantum fluctuations in curved space and the theory of gravitation'',
  Sov.\ Phys.\ Dokl.\  {\bf 12 } (1968)  1040-1041.
  
  
  \bibitem{Visser:Sakharov}
  M.~Visser,
  ``Sakharov's induced gravity: A modern perspective'',
  Mod.\ Phys.\ Lett.\  {\bf A17 } (2002)  977-992.
  [gr-qc/0204062].

   
   \bibitem{Barcelo}
  C.~Barcel\'o, S.~Liberati, M.~Visser,
  ``Analogue gravity'',
  Living Rev.\ Rel.\  {\bf 8 } (2005)  12.
  Updated and republished as  Living Rev.\ Rel.\  {\bf 14 } (2011)  3.
  [gr-qc/0505065].

 
   \bibitem{Visser:ergo}
  M.~Visser,
  ``Acoustic black holes: Horizons, ergospheres, and Hawking radiation'',
  Class.\ Quant.\ Grav.\  {\bf 15 } (1998)  1767-1791.
  [gr-qc/9712010].

    

\bibitem{Sotiriou:2006}
  T.~P.~Sotiriou, S.~Liberati,
  ``Field equations from a surface term'',
  Phys.\ Rev.\  {\bf D74 } (2006)  044016.
  [gr-qc/0603096].
  
  \bibitem{Sotiriou:2007}
  T.~P.~Sotiriou, S.~Liberati,
  ``Reply to ``Can gravitational dynamics be obtained by diffeomorphism invariance of action?'''',
  Phys.\ Rev.\  {\bf D75 } (2007)  068502.
  [gr-qc/0703080].


  \bibitem{Liberati:2009}
  S.~Liberati, F.~Girelli, L.~Sindoni,
  ``Analogue models for emergent gravity'',
  [arXiv:0909.3834 [gr-qc]].
  
  \bibitem{Sindoni:2009}
  L.~Sindoni, F.~Girelli, S.~Liberati,
  ``Emergent gravitational dynamics in Bose-Einstein condensates'',
  [arXiv:0909.5391 [gr-qc]].
  
  \bibitem{Sindoni:2010}
  L.~Sindoni,
  ``Emergent gravitational dynamics from multi-BEC hydrodynamics?'',
  Phys.\ Rev.\  {\bf D83 } (2011)  024022.
  [arXiv:1011.4411 [gr-qc]].
  
  
  \bibitem{Chirco:2010}
  G.~Chirco, C.~Eling, S.~Liberati,
  ``Reversible and irreversible spacetime thermodynamics for general Brans-Dicke theories'',
  Phys.\ Rev.\  {\bf D83 } (2011)  024032.
  [arXiv:1011.1405 [gr-qc]].
  
    
   

 
  \bibitem{Verlinde}
  E.~P.~Verlinde,
  ``On the origin of gravity and the laws of Newton'',
  JHEP {\bf 1104 } (2011)  029.
  [arXiv:1001.0785 [hep-th]].\\
  
  \bibitem{Smolin}
  L.~Smolin,
  ``Newtonian gravity in loop quantum gravity'',
  [arXiv:1001.3668 [gr-qc]].
  
  \bibitem{Wang}
  T.~Wang,
  ``The Coulomb force as an entropic force'',
  Phys.\ Rev.\  {\bf D81 } (2010)  104045.
  [arXiv:1001.4965 [hep-th]].
  
  \bibitem{Freund}
  P.~G.~O.~Freund,
  ``Emergent gauge fields'',
  [arXiv:1008.4147 [hep-th]].
  
  \bibitem{Nicolini}
  P.~Nicolini,
  ``Entropic force, noncommutative gravity and un-gravity,''
  Phys.\ Rev.\  {\bf D82 } (2010)  044030.
  [arXiv:1005.2996 [gr-qc]].
  
  \bibitem{Cai}
  R.~-G.~Cai, L.~-M.~Cao, N.~Ohta,
  ``Friedmann equations from entropic force'',
  Phys.\ Rev.\  {\bf D81 } (2010)  061501.
  [arXiv:1001.3470 [hep-th]].
  
  \bibitem{Li}
  M.~Li, Y.~Wang,
  ``Quantum UV/IR relations and holographic dark energy from entropic force'',
  Phys.\ Lett.\  {\bf B687 } (2010)  243-247.
  [arXiv:1001.4466 [hep-th]].
  
  \bibitem{Easson}
  D.~A.~Easson, P.~H.~Frampton, G.~F.~Smoot,
  ``Entropic accelerating universe'',
  Phys.\ Lett.\  {\bf B696 } (2011)  273-277.
  [arXiv:1002.4278 [hep-th]].
  
  \bibitem{Cai:2005}
  R.~-G.~Cai, S.~P.~Kim,
  ``First law of thermodynamics and Friedmann equations of Friedmann-Robertson-Walker universe,''
  JHEP {\bf 0502 } (2005)  050.
  [hep-th/0501055].
  
  \bibitem{Cai:2007}
  R.~-G.~Cai,
  ``Thermodynamics of apparent horizon in brane world scenarios,''
  Prog.\ Theor.\ Phys.\ Suppl.\  {\bf 172 } (2008)  100-109.
  [arXiv:0712.2142 [hep-th]].
  
  

  
  \bibitem{Hossenfelder}
  S.~Hossenfelder,
  ``Comments on, and Comments on Comments on, Verlinde's paper `On the origin of gravity and the laws of Newton' '',
  [arXiv:1003.1015 [gr-qc]].
  
  \bibitem{Kobakhidze}
  A.~Kobakhidze,
  ``Gravity is not an entropic force'',
  Phys.\ Rev.\  {\bf D83 } (2011)  021502.
  [arXiv:1009.5414 [hep-th]].
  
  \bibitem{Gao}
  S.~Gao,
  ``Is gravity an entropic force?'',
  Entropy {\bf 13 } (2011)  936-948.
  [arXiv:1002.2668 [physics.gen-ph]].
  
  \bibitem{Hu:2010}
  B.~L.~Hu,
  ``Gravity and nonequilibrium thermodynamics of classical matter'',
  Int.\ J.\ Mod.\ Phys.\  {\bf D20 } (2011)  697-716.
  [arXiv:1010.5837 [gr-qc]].
  
  \bibitem{Kobakhidze:2011}
  A.~Kobakhidze,
  ``Once more: Gravity is not an entropic force'',
  arXiv:1108.4161 [hep-th].
  

  
\bibitem{Muller}
Ingo M\"uller, \emph{A history of thermodynamics: The doctrine of energy and entropy}, (Springer, New York, 2007), 
ISBN-10: 3540462260;  ISBN-13: 978-3540462262. See especially ``Kinetic theory of rubber'',  pages 111--117.
\bibitem{Smith}
S.~B.~Smith, L.~Finzi, C.~Bustamante,  ``Direct mechanical measurements of the elasticity of single DNA molecules by using magnetic beads'', 
Science {\bf 258} \#  5085 (1992) 1122--1126. doi:10.1126/science.1439819.


  
\bibitem{Ramsey}
Norman~F.~ Ramsey, 
``Thermodynamics and statistical mechanics at negative absolute temperatures''. 
Physical Review {\bf 103 }(1956) 20--28. doi:10.1103/PhysRev.103.20

\bibitem{Purcell-Pound}
Edward~M.~Purcell and  Robert~V.~Pound, 
``A nuclear spin system at negative temperature'', 
Physical Review {\bf81} (1951) 279--280. doi:10.1103/PhysRev.81.279.

\bibitem{Mosk}
A.~P.~Mosk, 
``Atomic gases at negative kinetic temperature'', 
Physical Review Letters {\bf95} (2005) 040403, arXiv:cond-mat/0501344,
 doi:10.1103/PhysRevLett.95.040403.
 
 \bibitem{Kardar}
Mehran~Kardar, 
\emph{Statistical Physics of Particles}, 
(Cambridge University Press, England, 2007).


\bibitem{Shannon}
Claude~E.~Shannon, 
``A mathematical theory of communication'', 
Bell System Technical Journal, {\bf 27} (July/October 1948) 379--423, 623--656.

\bibitem{Landauer}
R.~Landauer, 
``Irreversibility and heat generation in the computing process'', 
IBM J. Res. Develop. {\bf5}, (1961) 183--191.

\bibitem{Brillouin}
Leon~Brillouin, 
\emph{Science and Information Theory}, 
(Dover, New York, [1956, 1962] 2004). ISBN 0-486-43918-6.





\bibitem{Unruh}
  W.~G.~Unruh,
  ``Notes on black hole evaporation'',
  Phys.\ Rev.\  {\bf D14 } (1976)  870.
  
  \bibitem{Bekenstein}
  J.~D.~Bekenstein,
  ``A universal upper bound on the entropy to energy ratio for bounded systems'',
  Phys.\ Rev.\  {\bf D23 } (1981)  287.
  
  \bibitem{Bekenstein:1973}
  J.~D.~Bekenstein,
  ``Black holes and entropy'',
  Phys.\ Rev.\  {\bf D7 } (1973)  2333-2346.
  
  \bibitem{Mechanix}
  J.~M.~Bardeen, B.~Carter, S.~W.~Hawking,
  ``The Four laws of black hole mechanics'',
  Commun.\ Math.\ Phys.\  {\bf 31 } (1973)  161-170.
  
  \bibitem{Hawking:1974}
  S.~W.~Hawking,
  ``Black hole explosions'',
  Nature {\bf 248 } (1974)  30-31.
  
 \bibitem{Hawking:1975}
  S.~W.~Hawking,
  ``Particle creation by black holes'',
  Commun.\ Math.\ Phys.\  {\bf 43 } (1975)  199-220. 
  



\end{thebibliography}
\end{document}